\documentstyle[12pt]{article}        
\input{psfig}
\newlength{\dinwidth}                       
\newlength{\dinmargin}                      
\def\lsim{\mathrel{\rlap{\lower4pt\hbox{\hskip1pt$\sim$}}
    \raise1pt\hbox{$<$}}}                
\def\gsim{\mathrel{\rlap{\lower4pt\hbox{\hskip1pt$\sim$}}
    \raise1pt\hbox{$>$}}}                
\begin{document}
\hbox to \hsize{
\hfill$\vtop{   \hbox{\bf MADPH-96-961\hspace{1cm}}
                \hbox{\bf TTP96-41 }
                \hbox{\bf hep-ph/9609274\\}
                \hbox{September 1996}}$ }
\begin{center}  \begin{Large} \begin{bf}
Jet Production in Deep Inelastic Scattering at Next-to-Leading Order\\
  \end{bf}  \end{Large}
  \vspace*{5mm}
  \begin{large}
Erwin Mirkes$^a$ and Dieter Zeppenfeld$^b$\\ 
  \end{large}
\end{center}
$^a$Institut f\"ur Theoretische Teilchenphysik, 
         Universit\"at Karlsruhe, D-76128 Karlsruhe, Germany\\
$^b$Department of Physics, University of Wisconsin, Madison, WI 53706, USA\\
\begin{quotation}
\noindent
{\bf Abstract:}
NLO corrections to jet cross sections in DIS at HERA are studied, with 
particular emphasis on the two jet final state.
High jet transverse momenta are a good criterion for the 
applicability of fixed order perturbation theory.
A ``natural'' scale choice is the average  $k_T^B$
of the jets in the Breit frame, which suggest analyzing the data
in different $<k_T^B>$ intervalls. 
\end{quotation}

An important topic to be studied at HERA is the 
production of multi-jet events in DIS,
where the expected good event statistics~\cite{exp} allows for 
precision tests of QCD~\cite{rom_exp}.  
Such tests require next-to-leading order (NLO) QCD corrections.
Full NLO corrections for one and two-jet production cross sections
and distributions are now available
and implemented in the fully differential
$ep \rightarrow n$ jets event generator MEPJET \cite{mz1},
which allows to analyze 
arbitrary jet definition schemes and 
general cuts in terms of parton 4-momenta. 
A variety of topics can be studied with these tools. 
They include:  
a) The determination of $\alpha_s(\mu_R)$ from dijet production 
over a range of scales, $\mu_R$,
b) The measurement of the 
gluon density in the proton (via $\gamma g\to q\bar q$),
c) Associated forward jet production in the low $x$ 
regime as a signal of BFKL dynamics \cite{mz_ws}.

The effects of NLO 
corrections and recombination scheme dependences 
on the 2-jet cross section were discussed in Refs.~\cite{mz1,mz2,mz3}
already for four different jet algorithms 
(cone, $k_T$, JADE, W) . While these effects are small in the cone
and $k_T$ schemes, very large corrections can 
appear in the $W$-scheme or the modified JADE scheme, which was introduced
for DIS in Ref.~\cite{wscheme}. 

At leading order (LO) the $W$ and the JADE scheme are equivalent. 
The NLO cross sections in the two schemes, however, can differ by 
almost a factor of two~\cite{mz1,mz2}, depending on the recombination scheme
and on the definition of the jet 
resolution mass ($(M_{ij}^2=(p_i+p_j)^2$ in the $W$ scheme versus
$M_{ij}^2=2E_iE_j(1-\cos\theta_{ij})$ defined in the lab frame 
in the JADE scheme). Trefzger and Rosenbauer~\cite{rom_exp}
find similarly large differences in the experimental
jet cross sections (which are in good agreement with 
MEPJET predictions\footnote{The  two jet rate in the $W$ scheme
(with $E$ recombination) and in the JADE scheme for corrected
ZEUS data are $18.6\pm 0.7 \%$ and $8.6\pm 0.5 \%$, respectively.
The corresponding 
NLO predictions from MEPJET for the same kinematics and the same
jet definitions  are $17.9 \% $ and 8.6 \%
(see T.~Trefzger \cite{rom_exp}).}), 
when the data are processed with  
exactly the same jet resolution mass
and recombination prescription as used in the theoretical calculation.
The large differences between and within the JADE and $W$ schemes
are caused by sizable single jet masses (compared to their energy), 
predominantly for jets in the central part of the detector.
Such single jet mass effects first appear in a NLO calculation where
a jet may be composed of two partons.
{\it Clearly, theoretical calculations must be matched to
experimental definitions and such potentially
large single jet mass effects must be taken into account}.

Previous programs \cite{projet,disjet} were limited to a $W$ type 
algorithm\footnote{DISJET \cite{disjet} and PROJET \cite{projet}
 are largely based on the
fact that the calculation of the jet resolution mass squared,
$M_{ij}^2$, can be done in a
lorentz invariant way, {\it i.e.} as in the $W$ scheme.
Only in LO does this agree with the JADE
definition, defined in the lab frame.}
 and are not flexible enough to take into account
the effects of single jet masses or differences between recombination schemes. 
In addition, approximations were made 
to the matrix elements in these programs which are not valid in large regions
of phase space~\cite{mz1}.
These problems are reflected in inconsistent values for $\alpha_s(M_Z^2)$ 
[ranging, for example, from 0.114 to 0.127 in the H1 analysis \cite{exp},
(see K. Rosenbauer \cite{rom_exp})], when these programs are used to analyze
the data with different recombination schemes. Because of these problems,
the older programs cannot be used for precision studies at NLO in their 
present form~\cite{kosower}. In order to reduce theoretical errors,
previous analyses \cite{exp} should be repeated with 
MEPJET or a similar flexible Monte Carlo program~\cite{disent}.
A first reanalysis, with MEPJET, of H1 data by K. Rosenbauer
yields a markedly lower central value, $\alpha_s(M_Z^2)=0.112$,
which is {\it independent} of the recombination scheme (used in both 
data and theory), and the $\alpha_s(\mu_R^2)$ extracted from 
different kinematical bins follows nicely the expectation from the
renormalization group equation. 
A similar reanalysis of the ZEUS data has already been performed 
by T. Trefzger, also with MEPJET. 

Single jet mass effects and  recombination scheme dependences
are fairly small in the cone and $k_T$ schemes \cite{mz1} which, therefore,
appear better suited for precision
QCD tests. In the following, we concentrate on these two and 
the $E$ recombination scheme.
A first issue which must be addressed is the dependence of the NLO
2-jet cross section on the renormalization scale, $\mu_R$, and the 
factorization scale, $\mu_F$. The chosen scale should be characteristic 
for the QCD portion of the process at hand. For dijet invariant 
masses, $m_{jj}$, below $Q$ we are in the DIS limit and $Q$ 
is expected to be the relevant scale. For large dijet invariant masses, 
however, $m_{jj}\gg Q$, the situation is more like in dijet production at 
hadron colliders and the jet transverse momenta set the physical scale of the 
process. A variable which interpolates between these two limits is the sum of 
jet $k_T$s in the Breit frame~\cite{mz3}, $\sum_j \,k_T^B(j)$.
Here, $(k_T^{B}(j))^2$ is defined by $2\,E_j^2(1-\cos\theta_{jP})$, where 
the subscripts $j$ and $P$ denote the jet and proton, respectively.
$\sum_j \,k_T^B(j)$ approaches $Q$ in the parton limit and it corresponds to 
the sum of jet transverse momenta, $p_T^B$, (with respect to 
the $\gamma^*$-proton direction) when the photon virtuality
 becomes negligible. We use this ``natural'' scale for multi-jet 
production in DIS in the following.

A good measure of the improvement of a  NLO over a LO prediction
is provided by the residual scale dependence of the cross section. 
As an example we use the $k_T$ algorithm (implemented in the Breit frame)
as described in Ref.~\cite{kt}. One finds 
very similar results for the cone scheme.
Kinematical cuts are imposed on the final state lepton and jets
to closely model the H1 event 
selection~\cite{which}. More specifically, we require 
10~GeV$^2<Q^2<10000$ GeV$^2$, $0.01 < y < 1$, $0.0001<x<1$,
and an energy cut of $E(l^\prime)>10$~GeV and a cut on the 
pseudo-rapidity $\eta=-\ln\tan(\theta/2)$
of the scattered lepton. This $\eta$ cut is $Q^2$ dependent:
$ -2.794 < \eta(l^\prime)<-1.735$\, for $Q^2 < 100$ GeV$^2$
and $ -1.317 < \eta(l^\prime)<2.436$ \,for $Q^2 > 100$ GeV$^2$.
In addition, we require $ -1.154 < \eta(j)< 2.436$.
The hard scattering scale, $E_T^2$, in the $k_T$ algorithm is
fixed to 40 GeV$^2$ and $y_{cut}=1$ is the resolution parameter 
for resolving the macro-jets. 

\begin{figure}[htb]
\psfig{figure=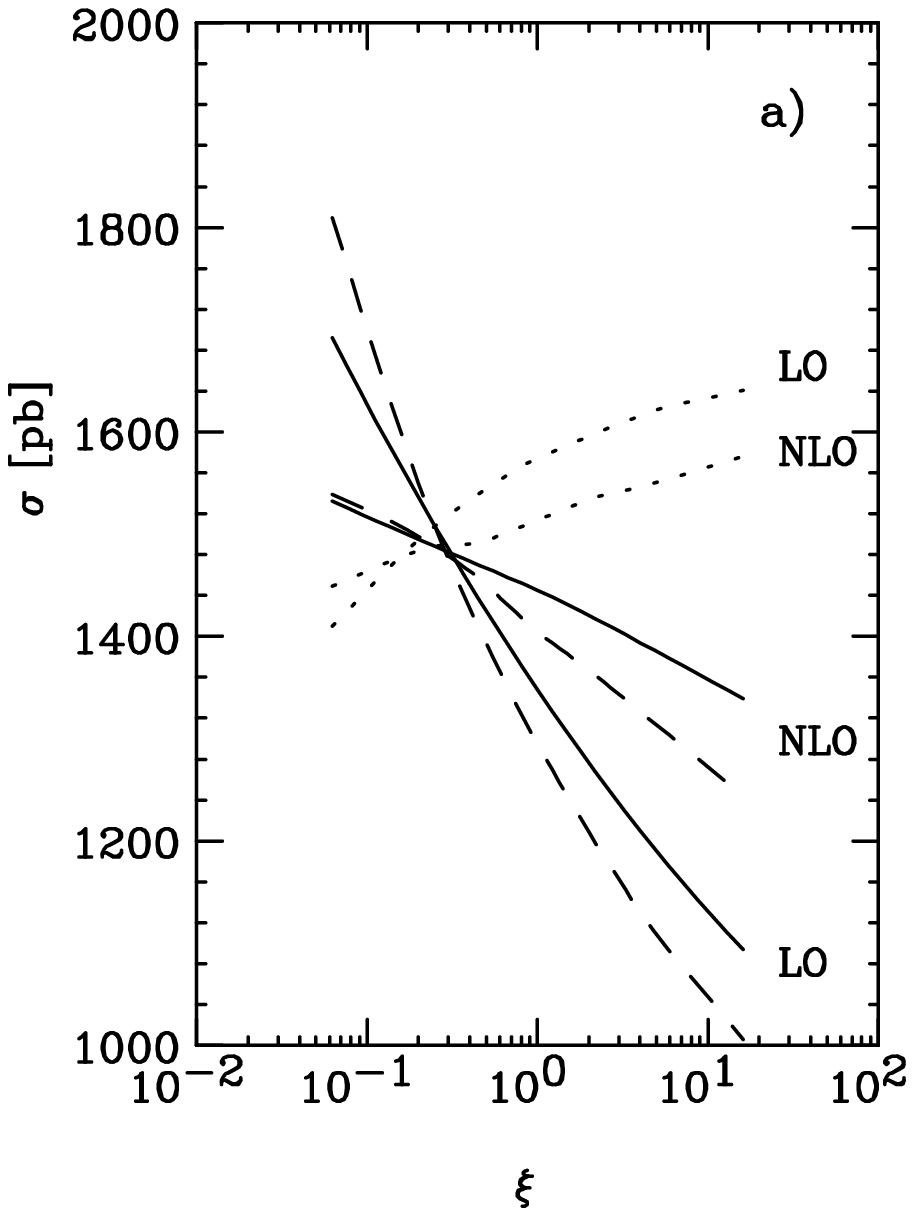,height=2.7in}
\vspace{-6.9cm}
\hspace*{5.5cm}
\psfig{figure=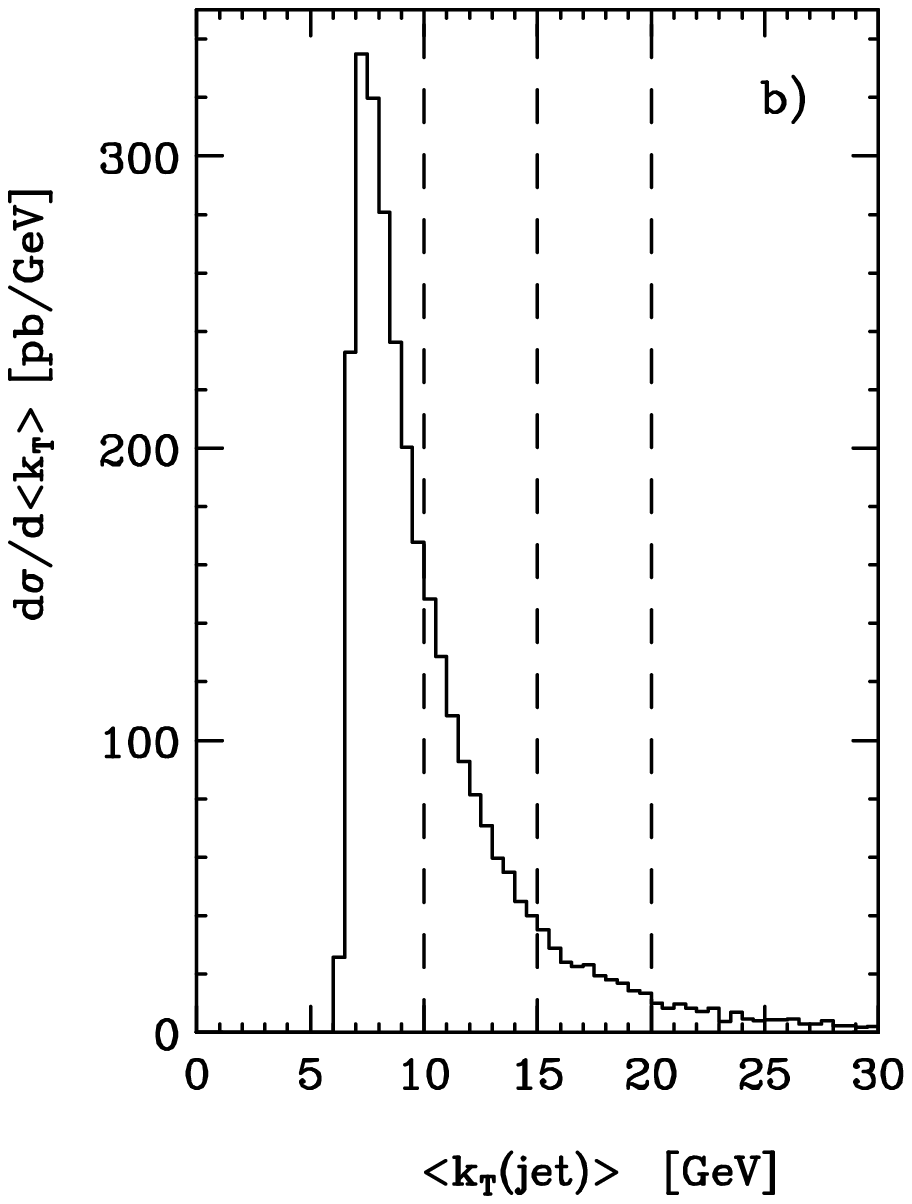,height=2.7in}
\vspace{-6.3cm}
\vspace{6.3cm}
\psfig{figure=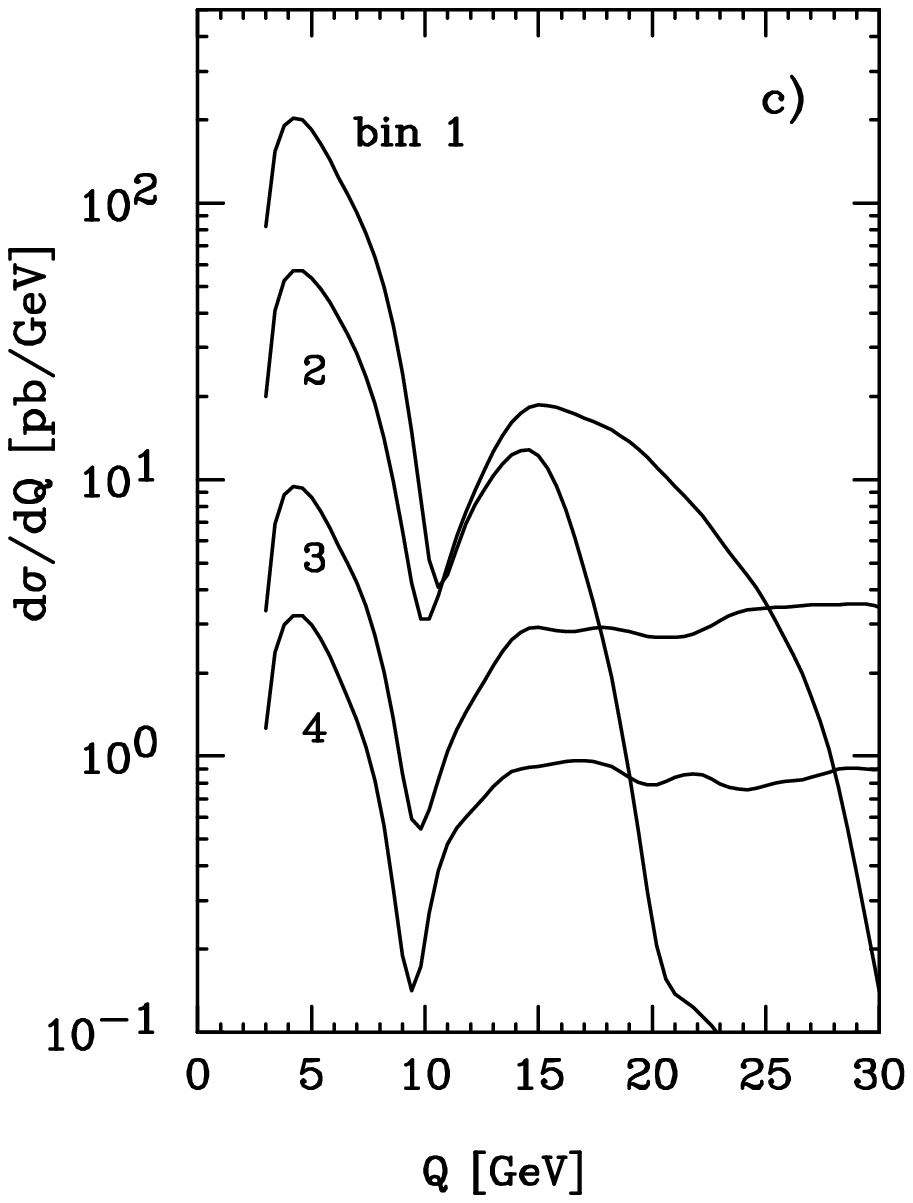,height=2.7in}
\caption{
a) Dependence of the two-jet exclusive cross 
section in the $k_T$ scheme 
on the  scale factor $\xi$.
The solid curves are for $\mu_R^2=\mu_F^2=\xi\;
(\sum_i\;k_T^B(i))^2$, while for
the dashed (dotted) curves only 
$\xi_R=\xi$ ($\xi_F=\xi$)  is varied but $\xi_F=1/4$ ($\xi_R=1/4$) is fixed.
Results are shown 
for the LO and NLO calculations. 
b) NLO  $<k_T^B>$ distribution for the two-jet exclusive cross section.
c) NLO $Q$  distribution for the four bins in b).
} 
\label{fig1}
\end{figure}

Fig.~1a shows the scale dependence of the dijet cross section in
LO and NLO  for the $k_T$ scheme.
The LO (NLO) results are based on 
the LO (NLO) parton distributions of GRV \cite{grv} 
together with
the one-loop (two-loop) formula with five flavors
for the strong coupling constant.
The scale factors $\xi$  are defined via
\begin{equation}
 \mu_R^2 = \xi_R\;(\sum_i \,k_T^B(i))^2\, ,
\hspace{1cm}
 \mu_F^2 = \xi_F\;(\sum_i \,k_T^B(i))^2\,.
\label{xidef}
\end{equation}
The LO variation by a factor 1.55 is reduced to a 11\%
variation at NLO when both scales are varied simultaneously 
over the plotted range (solid curves).
Also shown is the $\xi=\xi_R$ dependence of LO and NLO cross
sections at fixed $\xi_F=1/4$ (dashed curves)
and the $\xi=\xi_F$ dependence of LO and NLO cross
sections at fixed $\xi_R=1/4$ (dotted curves).
The NLO corrections substantially reduce the
renormalization and factorization scale dependence.
If not stated otherwise, we fix the scale factors to
$\xi=\xi_R=\xi_F=1/4$ in the following discussion.

Let us denote the average  $k_T^B$
of the (two) jets in the Breit frame by 
\begin{equation}
<k_T^B>=\frac{1}{2}\,\, (\sum_{j=1,2} \,k_T^B(j)).
\end{equation}
Fig.~1b shows the $<k_T^B>$ distribution for the NLO 2-jet exclusive 
cross section in the $k_T$ scheme. We divide the distribution into 
four $<k_T^B>$ bins (suggesting a separate determination 
of $\alpha_s(<k_T^B>^2)$ for each). 
The dependence of the NLO cross section on the 
scale factor, $\xi$, is shown in Table~1 for individual bins, and is 
typically below $\pm 5\%$. These fairly small theoretical
uncertainties  in the $k_T$ algorithm are due to the relatively
high value of the hard scattering scale, $E_T^2>40$ GeV$^2$
(or roughly equivalent cuts of $p_T^{lab},p_T^B \gsim 5$ GeV on the jets
in the cone scheme). Thus a precise measurement of $\alpha_s(<k_T^B>^2)$
should be possible.

\begin{table}
\caption{NLO (LO) 2-jet exlusive cross sections in pb for the four $<k_T^B>$ 
bins and their sum. Results are shown for three different choices of the scale
factor $\xi=\xi_R=\xi_F$.
}\label{table1}
\vspace{2mm}
\begin{tabular}{l|ccc}
        \hspace{0.8cm}
     &\hspace{5mm}  \mbox{$\xi=1   $  } \hspace{5mm}
     &\hspace{5mm}  \mbox{$\xi=1/4 $  } \hspace{5mm}
     &\hspace{5mm}  \mbox{$\xi=1/16$  } \hspace{5mm}\\
\hline\\[-3mm]
\mbox{bin 1:}\,\,\,\,\, 5~GeV  $<\,\,\,\,<k_T^B> \,\,\,\,<$ 10~GeV   
                         & 881 (821)    &  900 (907)  & 934 (999)   \\
\mbox{bin 2:}\,\, 10~GeV $<\,\,\,\,<k_T^B> \,\,\,\,<$ 15~GeV   
                         & 396 (357)    &  415 (403)  & 433 (461)  \\
\mbox{bin 3:}\,\, 15~GeV $<\,\,\,\,<k_T^B> \,\,\,\,<$ 20~GeV   
                         & 105 (102)    &  107 (118)  & 106 (137)    \\
\mbox{bin 4:}\,\, 20~GeV $<\,\,\,\,<k_T^B>$  
                         & 63   (68)     &   64  (80)   & 57 (95)    \\
\hspace{2cm} \mbox{sum of bins}    
                         & 1445 (1348)  & 1486 (1508) & 1530 (1692)  \\
\end{tabular}
\end{table}

The $Q$ distributions for the NLO exclusive dijet
cross section for these four bins in Fig.~1c show that even events 
with very large $<k_T^B>$ are dominated by the small $Q^2$ region.
(The dips in the $Q$ distribution around $Q=10$ GeV are a consequence
of the rapidity  cuts on the scattered lepton, see above). Thus there is
a qualitative difference betwen scale choices tied to $<k_T^B>$ versus
scales related to $Q$. One finds that $\mu_R^2,\;\mu_F^2=\xi Q^2$ gives
a much larger $\xi$ dependence for dijet events at NLO than the ones 
exhibited in Fig.~1a \cite{mz1}. This is the reason why scales tied 
to $k_T^B$ are better suited for QCD analyses of multijet 
events in DIS.

\end{document}